# Temporal logic with predicate abstraction.

Alexei Lisitsa and Igor Potapov *

Department of Computer Science,
University of Liverpool, Chadwick Building,
Peach St, Liverpool L69 7ZF, U.K.,
{alexei,igor}@csc.liv.ac.uk

**Abstract.** A predicate linear temporal logic $LTL_{\lambda=}$ without quantifiers but with predicate abstraction mechanism and equality is considered. The models of $LTL_{\lambda=}$ can be naturally seen as the systems of pebbles (flexible constants) moving over the elements of some (possibly infinite) domain. This allows to use $LTL_{\lambda=}$ for the specification of dynamic systems using some resources, such as processes using memory locations, mobile agents occupying some sites, etc. On the other hand we show that $LTL_{\lambda=}$ is not recursively axiomatizable and, therefore, fully automated verification of $LTL_{\lambda=}$ specifications is not, in general, possible.

## 1 Introduction

In this paper we consider a predicate linear temporal logic $LTL_{\lambda=}$ without quantifiers but with predicate abstraction mechanism. The idea of predicate abstraction goes back to M.Fitting who has proposed this as the general technique for obtaining the modal logics, which are, in a sense, intermediate between propositional and first-order. He suggested to extend a modal propositional logic $L$ by adding relation symbols, flexible constants and the operator of predicate abstraction, but no quantifiers. The abstraction is used as a scoping mechanism. Simple example of what the abstraction can be used for is given by the following two formulae: $\Diamond\langle\lambda x.P(x)\rangle(c)$ and $\langle\lambda x.\Diamond P(x)\rangle(c)$. The first one says that $P$ holds of what $c$ designates in alternative world, while the second one says that at an alternative world $P$ holds of what $c$ designates in a current world.

Such an extension $L_{\lambda(=)}$ (both with and without equality) can be alternatively seen as very restricted fragment of corresponding first-order variant $QL$ of $L$. It is proved in [3] that such extension when applied to $S5$ leads to the undecidable logic $S5_{\lambda=}$ but for many other classical modal logics $L$ their extensions $L_{\lambda(=)}$ are still decidable.

We apply such an extension to the classical propositional linear time logic. The models of $LTL_{\lambda=}$ can be naturally seen as the systems of pebbles (flexible constants) moving over the elements of some (possibly infinite) domain. This provides an abstract view on dynamic systems using some resources, such as processes using memory locations, mobile agents occupying some sites, etc. Thus, despite being very restricted extension of propositional temporal logic, $LTL_{\lambda=}$ is suitable for specification of such systems. However, we show, as a main result of this paper, that $LTL_{\lambda=}$ is not only undecidable, but even is not recursively axiomatizable. It follows that automatic verification of $LTL_{\lambda=}$ specifications is not, in general, possible.

---

* Work partially supported by NAL/00684/G NF grant.

The paper is organised as follows. In the next section we present a syntax and semantics of $LTL_{\lambda=}$ . In Section 3 we demonstrate the expressive power of $LTL_{\lambda=}$ by giving a range of examples of properties expressible in $LTL_{\lambda=}$ , show the limitations of $LTL_{\lambda=}$ and discuss possible applications of $LTL_{\lambda=}$ for specifications of protocols. In Section 4 we present main ideas of modelling counter machines by pebble systems. In Section 5 we use these ideas for modelling Minsky machines computations in $LTL_{\lambda=}$ and prove the main result. We conclude the paper by Section 6.

## 2 Syntax and Semantics

The content of this section is an adaptation of the corresponding section of [3] to the case of temporal logic.

Let $\mathcal{V} = \{x, y, x_1, \ldots\}$ is an alphabet of *variables* and $\mathcal{C} = \{a, b, c, c_1, \ldots\}$ is an alphabet of *constant symbols*. For each $n$ let $\mathcal{R}^n = \{P, R, R_1, \ldots\}$ is an alphabet of $n$-ary relational symbols. We refer to the tuple $\mathcal{L} = \langle \mathcal{V}, \mathcal{C}, \mathcal{R} \rangle$ as to the alphabet.

One may include or not an equality $=$ in the alphabet of binary relations symbols. In this paper we consider only the case with equality. A *term* is a constant symbol or a variable.

**Definition 1.** The set of $LTL_{\lambda=}$ -formulas (in the alphabet $\mathcal{L}$) and their free variables, are defined as follows.

1. If $R$ is an $n$-ary relation symbol and $x_1, x_2, \ldots$ are variables, then $R(x_1, x_2, \ldots x_n)$ is a formula with $x_1, x_2, \ldots$ as its free variable occurrences.
2. if $\varphi$ is a formula, then $\neg\varphi$, $\Box\varphi, \Diamond\varphi, \bigcirc\varphi, \blacksquare\varphi, \blacklozenge\varphi, \odot\varphi$ are formulas. Free variable occurrences are those of $\varphi$
3. If $\varphi$ and $\psi$ are formulas, then $(\varphi \land \psi)$, $(\varphi \lor \psi)$ and $(\varphi \to \psi)$ are formulas. Free variable occurrences are those of $\varphi$ together with those of $\psi$.
4. If $\varphi$ is a formula, $x$ is a variable, and $t$ is a term, then $\langle \lambda x.\varphi \rangle(t)$ is a formula. Free variable occurences are those of $\varphi$, except for occurrences of $x$, together with $t$ if it is a variable.

A formula $\varphi$ is called a *sentence* iff it does not have free variable occurences.

Formulae of $LTL_{\lambda=}$ are interpreted in first-order temporal models with the time flow isomorphic to the structure $\langle \mathbb{N}, \leq, succ \rangle$, where $\mathbb{N}$ is the set of natural numbers, $\leq$ is usual order relation on $\mathbb{N}$, and $succ$ is a successor operation on $\mathbb{N}$.

**Definition 2.** A *model* is a structure $\mathfrak{M} = \langle D, \mathcal{I} \rangle$ , where:

1. $D$ is a non-empty set, the domain;
2. $\mathcal{I}$ is an interpretation mapping that assigns:
   – to each constant symbol some function from $\mathbb{N}$ to $D$;
   – to each $n$-ary relation symbol some function from $\mathbb{N}$ to $2^{D^n}$ (the power set of $D^n$)

3. $\mathcal{I}(=)$ is the constant function assigning the equality relation on $D$ to every moment of time from $\mathbb{N}$.

First-order (non-temporal) structures corresponding to each point of time will be denoted $\mathfrak{M}_n = \langle D, \mathcal{I}(n) \rangle$. Intuitively, the interpretations of $LTL_{\lambda=}$ -formulae are sequences of first-order structures, or *states* of $\mathfrak{M}$, such as $\mathfrak{M}_0, \mathfrak{M}_1, \ldots, \mathfrak{M}_n \ldots$.

An *assignment* in $D$ is a function $\mathfrak{a}$ from the set $\mathcal{V}$ of individual variables to $D$. Thus we assume that (individual) variables of $LTL_{\lambda=}$ are *rigid*, that is assignments do not depend on the state in which variables are evaluated. In contrast, as follows from Definition 2 the constants are assumed to be non-rigid (flexible), that is their interpretations depend on moments of time (states). For a constant $c$ we call element $I(c)(n)$ of $D$ also a *designation* of $c$ at the moment $n$. The assignment $\mathfrak{a}$ is extended to the assignment $(\mathfrak{a} * \mathcal{I}) : (\mathcal{C} \cup \mathcal{V}) \to D$ of all terms in the usual way:

1. For a variable $x$, $(\mathfrak{a} * \mathcal{I})(x, n) = \mathfrak{a}(x)$;
2. For a constant $c$, $(\mathfrak{a} * \mathcal{I})(c, n) = \mathcal{I}(c)(n)$.

If $P$ is a predicate symbol then $P^{\mathcal{I}(n)}$ (or simply $P^n$ if $\mathcal{I}$ is understood) is the interpretation of $P$ in the state $\mathfrak{M}_n$.

**Definition 3.** The *truth-relation* $\mathfrak{M}_n \models^{\mathfrak{a}} \varphi$ (or simply $n \models^{\mathfrak{a}} \varphi$, if $\mathfrak{M}$ is understood) in the structure $\mathfrak{M}$ for the assignment $\mathfrak{a}$ is defined inductively in the usual way under the following semantics of temporal operators:

$n \models^{\mathfrak{a}} \bigcirc \varphi$ iff $n + 1 \models^{\mathfrak{a}} \varphi$;
$n \models^{\mathfrak{a}} \Diamond \varphi$ iff there is $m \geq n$ such that $m \models^{\mathfrak{a}} \varphi$;
$n \models^{\mathfrak{a}} \Box \varphi$ iff $m \models^{\mathfrak{a}} \varphi$ for all $m \geq n$;
$n \models^{\mathfrak{a}} \odot \varphi$ iff $n - 1 \models^{\mathfrak{a}} \varphi$;
$n \models^{\mathfrak{a}} \blacklozenge \varphi$ iff there is $0 \leq m \leq n$ such that $m \models^{\mathfrak{a}} \varphi$;
$n \models^{\mathfrak{a}} \blacksquare \varphi$ iff $m \models^{\mathfrak{a}} \varphi$ for all $0 \leq m \leq n$.

For the case of abstraction we have:

$n \models^{\mathfrak{a}} \langle \lambda x. \varphi \rangle(t)$ iff $n \models^{\mathfrak{a}'} \varphi$, where $\mathfrak{a}'$ coincide with $\mathfrak{a}$ on all variables except $x$ and $\mathfrak{a}'(x) = (\mathfrak{a} * \mathcal{I})(t, n)$.

A formula $\varphi$ is said to be *satisfiable* if there is a first-order structure $\mathfrak{M}$ and an assignment $\mathfrak{a}$ such that $\mathfrak{M}_0 \models^{\mathfrak{a}} \varphi$. If $\mathfrak{M}_0 \models^{\mathfrak{a}} \varphi$ for every structure $\mathfrak{M}$ and for all assignments $\mathfrak{a}$ then $\varphi$ is said to be *valid*. Note that formulae here are interpreted in the initial state $\mathfrak{M}_0$.

We conclude this section by introducing an useful notation. Given a model $\mathfrak{M} = \langle D, \mathcal{I} \rangle$ and constant $c$ denote by $V_c^n$ the set of elements of $D$ visited by $c$ up to the moment $n$, that is $V_c^n = \{\mathcal{I}(m)(c) \mid 0 \leq m \leq n\}$. Then $V_c = \cup_{i \in \mathbb{N}} V_c^i$ is the set of elements of $D$ visited by $c$ in the model $\mathfrak{M}$.

## 3 Properties expressible in $LTL_{\lambda=}$

Despite being very restricted fragment of the first-order temporal logic the logic $LTL_{\lambda=}$ can express many non-trivial properties of its models. Because of the interpretation given to flexible constants (by a function from $\mathbb{N}$ to $D$) they can be though of as the *pebbles* moving over elements of the domain as the time goes by. Then, even in the absence of any other predicate symbols except equality, one can express dynamic properties of the system of pebbles evolving in time:

- $AlwaysNew(d) \Leftrightarrow \Box\langle\lambda x. \bigcirc \Box\langle\lambda y.y \neq x\rangle(d)\rangle(d)$. The constant d has different designations at different moments of time (the pebble d occupies different positions in different moments of time).
- $Same(a,b) \Leftrightarrow \langle\lambda x.\langle\lambda y.x = y\rangle(a)\rangle(b)$. The constants a and b have the same designation now. (The pebbles are on the same place now).
- $SameInPast(a,d) \Leftrightarrow \langle\lambda x.\blacklozenge\langle\lambda y.y = x\rangle(d)\rangle(a)$. The constant a has the same designation as d had in the past.
- $NoChange(c) \Leftrightarrow \langle\lambda x. \bigcirc \langle\lambda y.x = y\rangle(c)\rangle(c)$. The constant $c$ has the same designation at the current and next moments of time.
- $AlwaysReturn(a) \Leftrightarrow \Box\langle\lambda x.\langle\bigcirc\Diamond\langle\lambda y.x = y\rangle(a)\rangle(a)$. The pebble a always return to the place it occupies at any given moment of time.

Next, we present two examples of more complex formulae, which will play special role later on:

- $NextNew1(a,d,c) \Leftrightarrow \langle\lambda w.\langle\lambda x.\blacklozenge(\langle\lambda y.y = x\rangle(d)\land\bigcirc\langle\lambda v.(v = w)\rangle(d))\rangle(a)\rangle(c)\land \langle\lambda z. \bigcirc \langle\lambda t.(t = z)\rangle(a)\rangle)(c)$
- $NextNew2(a,d) \Leftrightarrow \bigcirc\langle\lambda w.\odot\langle\lambda x.\blacklozenge(\langle\lambda y.y = x\rangle(d)\land\bigcirc\langle\lambda v.(v = w)\rangle(d))\rangle(a)\rangle(a)$

Both formulae express the same fact about the behaviour of pebbles $a$ and $d$: the pebble $a$ moves in the next moment of time to the position to which the pebble $d$ has moved from the position which $a$ is occupying now. In other words $a$ moves in the same way as $d$ did from the same position. The difference between two formulae is that NextNew1 does not use "last" operator $\odot$ but at the expense of an additional flexible constant.

In the above formulae no predicate symbols except equality was used. One example of the formula with additional binary predicate $E$ is

$$\Box(\langle\lambda x.\langle\lambda y.(E(x,y) \leftrightarrow (\Box E(x,y) \land \blacksquare E(x,y)))\rangle(c_1)\rangle(c_2))$$

This formula says that predicate $E$, restricted to the pairs of elements of domain, visited by first and second constant, respectively, is rigid. But $E$ may well have different interpretations at different moments of times on all other pairs of elements. This example illustrates also the *pebble locality* of properties expressible by $LTL_{\lambda=}$ formulae. Pebble locality of the property means the property depends only on the interpretations of predicate symbols on the elements of the domains ever visited by pebbles. In fact, only such properties are expressible in $LTL_{\lambda=}$ as the Proposition 1 shows.

**Definition 4.** Given an alphabet $\mathcal{L} = \langle\mathcal{V},\mathcal{C},\mathcal{R} = \cup_{n\in\mathbb{N}}\mathcal{R}^n\rangle$. Two models $\mathfrak{M} = \langle D,\mathcal{I}\rangle$ and $\mathfrak{M}' = \langle D',\mathcal{I}'\rangle$ are said to be *pebble equivalent* with respect to $\mathcal{L}$ and this is denoted by $\mathfrak{M} \equiv^p_{\mathcal{L}} \mathfrak{M}'$ iff

- $\forall c \in \mathcal{C} \ \forall i \in \mathbb{N} \ (\mathcal{I}(c)(i) \in D \cap D' \land \mathcal{I}'(c) \in D \cap D')$ (constants of both models are interpreted on the common part of the domains; actually, it follows from the next clause);
- $\forall c \in \mathcal{C} \ \forall i \in \mathbb{N} \ \mathcal{I}(c)(i) = \mathcal{I}'(c)(i)$ (interpretations of any constant at any moment of time coincide in both models);
- $\forall P \in \mathcal{R}^n \ \forall i \in \mathbb{N} \ \forall \bar{v} \in \underbrace{V \times V \times \ldots V}_{n \text{ times}} [\bar{v} \in \mathcal{I}(P)(i) \Leftrightarrow \bar{v} \in \mathcal{I}'(P)(i)]$ for every arity $n$; here $V = \cup_{c \in \mathcal{C}} V_c$ (interpretations of predicate symbols coincide in both models on the elements of the domains visited by constants).

**Proposition 1.** Let $\varphi \in LTL_{\lambda=}$ is a sentence in alphabet $\mathcal{L}$. Then it can not distinguish pebble equivalent models, that is $\mathfrak{M} \equiv^p_\mathcal{L} \mathfrak{M}' \Rightarrow [\mathfrak{M}_0 \models \varphi \Leftrightarrow \mathfrak{M}'_0 \models \varphi]$

**Proof.** The proof is based on easy induction on the formula structure using Definition 3. The induction assumption is that for any formula $\varphi$ possibly including free variables the truth value $n \models^a \varphi$ depends only on the assignment of free variables, on the interpretation of flexible constants and on the interpretation of predicate symbols on elements of $V$. Since a sentence does not inlude free variable occurences, the statement of the proposition follows. □

We use pebble locality now to show that $LTL_{\lambda=}$ is less experessive that the full first-order linear temporal logic $FOLTL$ where unrestricted quantification is allowed. In what follows we assume the semantics of $FOLTL$ with flexible predicates and rigid constants. Consider the $FOLTL$ formula in a vocabulary consisting single binary predicate symbol $E$:

$$\Box(\forall x \forall y E(x,y) \Leftrightarrow \bigcirc E(x,y))$$

The formula expresses persistence of the interpretation of $E$ along the time flow. Such a property can not be expressed by any formula of $LTL_{\lambda=}$ as the following proposition shows.

**Proposition 2.** *Let $\mathcal{L} = \langle \mathcal{V}, \mathcal{C}, \mathcal{R} \rangle$ be an alphabet with $\mathcal{R}$ containing a single binary predicate symbol, and $\mathcal{C}$ be an arbitrary set of flexible constant symbols. For any satisfiable sentence $\varphi$ of $LTL_{\lambda=}$ in alphabet $\mathcal{L}$ there is a model satisfying $\varphi$ with non persistent interpretation of $E$.*

**Proof.** Take any model $\mathfrak{M} = \langle D, \mathcal{I} \rangle$ such that $\mathfrak{M} \models \varphi$. Then construct a new model $\mathfrak{M}' = \langle D', \mathcal{I}' \rangle$ where $D' = D \cup D''$, $D'' \cap D = \emptyset$ and $D''$ contains two elements. For any constant $c$ put $\mathcal{I}'(c) = \mathcal{I}(c)$, i.e. constants do not visit any element of $D''$. Choose any interpretation $\mathcal{I}'(E)$ of $E$ which satisfies:

- $\forall x, y \in D \ \forall n \in \mathbb{N} \ (\langle x, y \rangle \in \mathcal{I}'(E)(n) \Leftrightarrow \langle x, y \rangle \in \mathcal{I}(E)(n))$
- $\forall x, y \in D'' \ \forall k \in \mathbb{N} \ (\langle x, y \rangle \in \mathcal{I}'(E)(2k) \land \langle x, y \rangle \notin \mathcal{I}'(E)(2k+1))$

Notice that we don't care how $\mathcal{I}'(E)$ is defined on all other pairs of elements.

Then we have $\mathfrak{M}'$ is pebble equivalent to $\mathfrak{M}$ and therefore $M' \models \varphi$. The interpretation of $E$ in $\mathfrak{M}'$ depends on time. □

### 3.1 Pebble systems and agents using resources

The above metaphor of pebble system may also be seen as the very abstract model of computational processes (agents) using some resources. In such a model a pebble, or, indeed non-rigid constant $c$ may be thought of as an computational process and elements of the domain as the abstract resources. Then, if at some moment of time a designation of $c$ is an element $x$ of the domain, one may understand it as *"c uses the resource $x$"*. To model the situation with processes, or agents using several resources at the same time, one may associate with an agent a *set* of flexible constants (pebbles). Another natural reading of the above situation may be *"the mobile agent c resides at the host $x$"*. Taking this point, the formulae of $LTL_{\lambda=}$ can be used to specify protocols, policies or requirements for agents operating within the common pool of resources. Pointing out this possibility, we restrict ourselves in this paper with the simple example of communicating protocol for mobile agents.

### 3.2 An example of communication protocol for Mobile Agents

Let us suppose the following scenario where a group of communicating mobile agents explore some hosts and transmit messages to each other. Because mobile agents can move autonomously from host to host, they cannot reliably know the location of their communication peer. Therefore, a practical communication protocol somehow must keep track of agent locations, allowing each agent to send messages to its peers without knowing where they physically reside.

There are many mobile agent tracking protocols, that use a forwarding pointers mechanism[1]. It means that each host on mobile agents migration path keeps a forwarding pointer to the next host on the path. The classical primitive for such protocol is based on *knowledge of each sender the target agent's home.* So messages are sent to the agent's home and forwarded to the target object along the forwarding pointers. Interesting alternative is a primitive *find a host, which was visited both by sender and receiver*[1]. Using this primitive the messages are again forwarded to the target object along the forwarding pointers but from the host where the mobile agents migration paths intersect (see Figure 1).

We can specify the use of this primitive in $LTL_{\lambda=}$ as follows. For simplicity we assume that receiver always either do not move or move to the new host (never revisiting the hosts it already visited).

Let flexible constants $s$ and $r$ denote communicating mobile agents (sender and receiver, respectively) and $m$ denotes the message. Then $LTL_{\lambda=}$ -formula

$$Same(s,m) \wedge \bigcirc (\lambda z.\langle\blacklozenge\langle\lambda x.\blacklozenge\langle\lambda y.x=y \wedge y=z\rangle(r)\rangle(s)\rangle(m)) \wedge \bigcirc \square NextNew2(m,r)$$

describes the above protocol: at some moment of time message $m$ is on the same hosts as $s$, then it moves along the path of $r$, starting from a host which both $s$ and $r$ have visited (and $r$ done it no later than $s$).

---

[1] We don't consider the issue of implementation of such a primitive and note that this can be done in various ways.

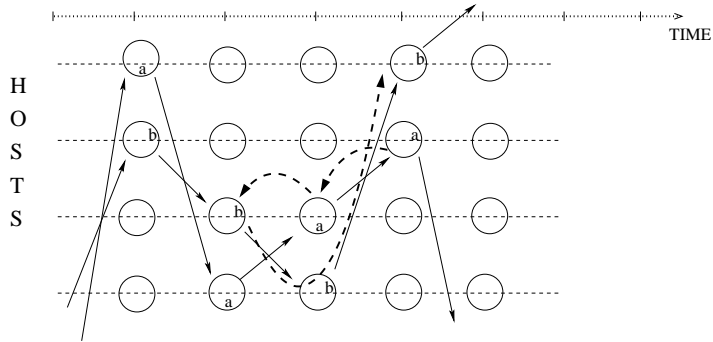

**Fig. 1.** A mobile agent tracking protocols: the operation of sending message from agent $a$ to agent $b$.

It should be clear now, that $LTL_{\lambda=}$ is expressive enough to formulate also the correctness conditions for such protocols, like *once the message sent, it will be delivered eventually to a receiver*. Of course, one needs to specify some extra conditions which would guarantee correctness: *receiver must stop and wait in order to receive a message* (otherwise the message may always be behind the receiver). Or, one may specify the different speed of messages and agents, which would guarantee delivery even to the agents "on move". One way of doing this in $LTL_{\lambda=}$ is to specify that messages can move to the new host every round (discrete moment of time), while mobile agents can move only every *second* round. Thus, the proof of correctness of the above protocol may be reduced to the validity checking for some $LTL_{\lambda=}$ -formulae. We don't pursue a goal of automatic verification of protocols via validity checking (theorem proving) for $LTL_{\lambda=}$ -formulae in this paper, but rather demonstrate a related negative result on $LTL_{\lambda=}$ itself: it is highly undecidable and, therefore, fully automated verification based on validity checking of $LTL_{\lambda=}$ -formulae is not possible.

## 4  Minsky machines and their modelling by pebbled sets

In this section we use a well known model of Minsky machine to show universality of pebbled sets model. Informally speaking, Minsky machine is a two counter machine that can increment and decrement counters by one and test them for zero. It is known that Minsky machines represents a universal model of computations [7]. Being of very simple structure the Minsky machine are very useful for proving undecidability results (see for example [4, 5]).

It is convenient to represent a counter machine as a simple imperative program $\mathcal{M}$ consisting of a sequence of instructions labelled by natural numbers from 1 to some $L$. Any instruction is one of the following forms:

$l$: ADD 1 to $S_k$; GOTO $l'$;
$l$: IF $S_k \neq 0$ THEN SUBTRACT 1 FROM $S_k$; GOTO $l'$ ELSE GOTO $l''$;
$l$: STOP.

where $k \in \{1, 2\}$ and $l, l', l'' \in \{1, \ldots, L\}$.

The machine $\mathcal{M}$ starts executing with some initial nonnegative integer values in counters $S_1$ and $S_2$ and the control at instruction labelled 1. We assume the semantics of all above instructions and of entire program is clear. Without loss of generality one can suppose that every machine contains exactly one instruction of the form $l$: STOP which is the last one ($l = L$). It should be clear that the execution process (run) is deterministic and has no failure. Any such process is either finished by the execution of $L$: STOP instruction or lasts forever.

As a consequence of the universality of such computational model the halting problem for Minsky machines is undecidable:

**Theorem 1 ([7]).** *It is undecidable whether a two-counter Minsky machine halts when both counters initially contain* 0.

We will use the following consequence of Theorem 1.

**Corollary 1.** *The set of all Minsky machines which begin with both counters containing* 0 *and do not halt is not recursively enumerable.*

Given any machine $\mathcal{M}$ (with initial values for the two counters) let us define its run $r^{\mathcal{M}}$ as a sequence of triples, or states of $r^{\mathcal{M}}$:

$$(l_1, p_1^0, p_2^0), (l_2, p_1^1, p_2^1), \ldots (l_{j+1}, p_1^j, p_2^j), \ldots$$

where $l_j$ is the label of the instruction to be executed at $j$th step of computation, $p_1^j$ and $p_2^j$ are the nonnegative integers within the first and the second counters, respectively, after completion of $j$th step of computation. Depending on whether $\mathcal{M}$ stops or not $r^{\mathcal{M}}$ can be finite or infinite.

Henceforth we will consider only the computations of the Minsky machines started with both counters containing 0. Thus we always put $p_1^0 = 0$, $p_2^0 = 0$ and $l_1 = 1$.

### 4.1 Modelling Minsky machines by systems of pebbles

We will show our main result on non-r.e. axiomatizability of $LTL_{\lambda=}$ by modelling the computations of two counter Minsky machines in that logic. In fact we are going to model such machines by pebble systems and then just express required properties of such systems in the logic. In this subsection we explain the main idea of modelling, leaving all details of $LTL_{\lambda=}$ representation to the next section. Actually we suggest two methods of Minsky Machine modelling using pebbles. Both of them are based on the same idea, but use different number of pebbles and basis operations, that require different temporal operators.

**Method 1.** Given a pebble system with tree pebbles $a$, $b$, $d$. We denote the set of all elements that was visited by a pebble $a$ ($b$) until the moment of time $i$ by $V_a^i$ ($V_b^i$). One may use then two pebbles, say $a$ and $b$ to model the counter's values as follows. We represent the counter's value at the moment $i$ as the *cardinality* of the set $V_a^i \cap \overline{V_b^i}$. Increasing one of the sets of elements visited by $a$, or by $b$ one may increase or decrease the counter value. Our modelling will ensure that $\forall i. V_b^i \subseteq V_a^i$. That means the counter's value at the moment $i$ is in fact $card(V_a^i) - card(V_b^i)$ (see Figure 2).

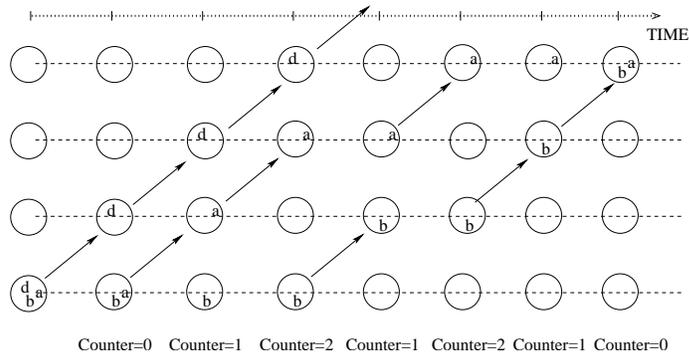

Counter=0  Counter=1  Counter=2  Counter=1  Counter=2  Counter=1  Counter=0

**Fig. 2.** First method of encoding

Due to peculiarities of logical representation we confine the range of the elements visited by both pebbles to the set of elements visited by another special pebble $d$. We require $d$ moves every time to the new element and we have $V_b^i \subseteq V_a^i \subseteq V_d^i$.

Let us show how to increase and decrease the cardinality of the set $V_a^i \cap \overline{V_b^i}$. Since the pebble $d$ generates unique sequence of elements from the domain as the time goes by we can use this unique sequence for increasing of the cardinality of $V_a^i$ or $V_b^i$ by one.

Let pebble $a$ ($b$) is on an element $x$ of the domain. Since $a$ is moving strictly along the path of $d$, the pebble $d$ has visited the element $x$ and moved to another element $y$. So in order to increase the cardinality of $V_a^i$ ($V_b^i$) by one we need to move the pebble $a$ ($b$) to the element $y$. In other words $a$ ($b$) moves in the same way as $d$ did from the same position.

We can increase (decrease) the value of counter by one or in other words increase (decrease) the cardinality of the set $C = V_a^i \cap \overline{V_b^i}$ by one if we increase the cardinality of the set $V_a^i$ ($V_b^i$) by one according to the above procedure. Since there is a strict order of unique elements that we use for moving pebbles along the path of $d$ we can easily test the emptiness of the counter or emptiness of the set $C$ by checking if the pebble $a$ and the pebble $b$ are on the same element (see Figure 3).

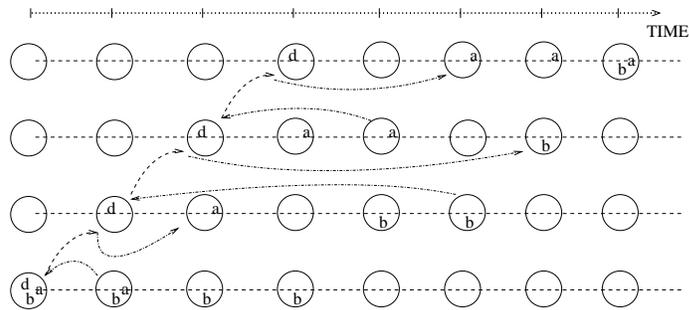

**Fig. 3.** Pebbels $a$ and $b$ moves in the same way as $d$ did from the same position.

**Method 2.** In the second method we again use pebbles $a$, $b$ and $d$, but it will be enough to model even two counters at the same time. We require $d$ to move to the new element of the domain every next time step. Also we are going to define the moving of pebbles $a$ and $b$ in such way that we have $V_a^i \subseteq V_d^i$ and $V_b^i \subseteq V_d^i$.

In contrast to the previous method we use here a different coding of counters. Let the pebble $a$ is on the element $u$ of the domain and the pebble $b$ is on the element $v$ of the domain and both of these elements have been already visited by $d$ in moments of time $i$ and $j$ correspondingly. In such case the cardinality of $V_d^i$ stands for one counter and $V_d^j$ stands for another one (see Figure 4).

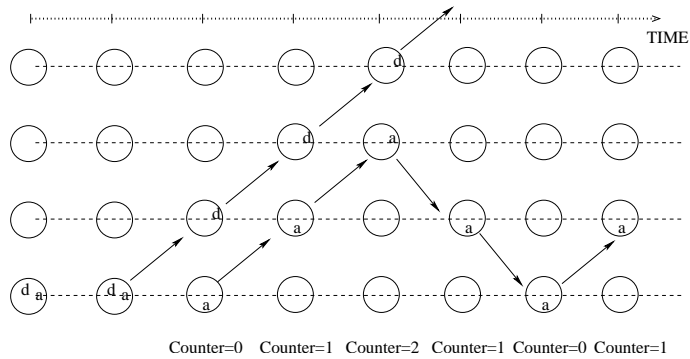

**Fig. 4.** Second method of encoding.

Now we define the increment operation, decrement operation and testing for zero in terms of pebbles $a$, $b$ and $d$. According to the above coding we can change first (second) counter by moving the pebble $a$ ($b$) along the path of the pebble $d$.

Let pebble $a$ is on a place $y$, which is an element of the domain and the pebble $d$ has visited element $y$ in the time step $i$ from some element $x$ and then moved to another element $z$ in time step $i+1$. So in order to increase the first counter we need to move the pebble $a$ to the element $z$ and for decreasing by one we need to move $a$ to the element $x$. Now the value of the counter will be represented by $V_d^{i+1}$ after increment operation and $V_d^{i-1}$ after decrement operation. In a similar way we can model another counter using the pebble $b$.

The final operation is testing for zero that can be modelled by checking if the pebble $a$ ($b$) is on the element which was an initial position of $d$.

## 5 Modelling of Minsky machines in $LTL_{\lambda=}$

In the translation of Minsky machines into formulae of $LTL_{\lambda=}$ we will use the formulae defined in the section 3. We implement the Method 1 from the previous section and note that Method 2 could be used instead.

### 5.1 Translation

Given a Minsky machine $\mathcal{M}$ defined by the sequence of instructions $c_1, \ldots c_L$ we define $LTL_{\lambda=}$ temporal formula $\chi^{\mathcal{M}}$ as follows.

Let $e_1, \ldots, e_L$ be flexible constants corresponding to instructions $c_1, \ldots, c_L$. Let $e_0$ and $f$ be two additional constants. The intention is to model the fact "$c_l$ is executed at the moment t" by coincidence of designations of $e_l$ and $f$ at the moment $t$. We denote by $Q_l$ the formula expressing this fact: $\langle \lambda x.\langle \lambda y.x = y\rangle(e_l)\rangle(f)$. Since we assume $c_L$ is the STOP instruction we will denote $Q_L$ alternatively as $Q_{stop}$. Further we have five more constants for modelling counters: $d, a_1, b_1, a_2, b_2$.

Then, for every instruction $c_l$, except $l$ : STOP, we define its translation $\chi(c_l)$ as follows:

**A.** <u>An instruction of the form</u>

$l$ : ADD 1 to $S_k$; GOTO $l'$;

is translated into the conjunction of the following formulae:

A1. $\Box(Q_l \to \text{NextNew}(a_k, d))$
A2. $\Box(Q_l \to \text{NoChange}(b_k))$
A3. $\Box(Q_l \to \text{NoChange}(a_{3-k}))$
A4. $\Box(Q_l \to \text{NoChange}(b_{3-k}))$
A5. $\Box(Q_l \to \bigcirc Q_{l'})$

Formulae A1–A4 ensure that in every temporal model $\mathfrak{M}$ for them, once we have $Q_l^n = $ **true** at a moment $n$, at the next moment the interpretation of the flexible constant $a_k$ changes to a new value, while $b_k, a_{3-k}$ and $b_{3-k}$ keep their interpretation intact. The formula A5 describes switching truth values of propositions $Q_i$ ($i \in \{1, \ldots L\}$) and the aim here is to model the transition from the instruction which is executed to the next one.

**B.** <u>An instruction of the form</u>

$l$ : IF $S_k \neq 0$ THEN SUBTRACT 1 FROM $S_k$; GOTO $l'$ ELSE GOTO $l''$;

is translated into the conjunction of the following formulae:

B1. $\Box(Q_l \land \neg\text{Same}(a_k, b_k) \to \text{NoChange}(a_k))$
B2. $\Box(Q_l \land \neg\text{Same}(a_k, b_k) \to \text{NextNew}(b_k, d))$
B3. $\Box(Q_l \land \neg\text{Same}(a_k, b_k) \to \text{NoChange}(a_{3-k}))$
B4. $\Box(Q_l \land \neg\text{Same}(a_k, b_k) \to \text{NoChange}(a_{3-k}))$
B5. $\Box(Q_l \land \text{Same}(a_k, b_k) \to \text{NoChange}(a_k) \land \text{NoChange}(b_k) \land \text{NoChange}(a_{3-k}) \land \text{NoChange}(b_{3-k}))$
B6. $\Box(Q_l \land \neg Same(a_k, b_k) \to \bigcirc Q_{l'})$
B7. $\Box(Q_l \land Same(a_k, b_k) \to \bigcirc Q_{l''})$

Formulae B1–B4 ensure that, in every temporal model for them, once we have $Q_l$ and the interpretations of $a_k$ and $b_k$ are different (meaning "$k$st counter has non-zero value") the interpretation of $b_k$ changes in the next moment of time, while interpretations of $a_k, a_{3-k}$ and $b_{3-k}$ still the same. Formula B5 ensures that, when $Q_l$ and interpretations of $a_k$ and $b_k$ are the same (meaning "counter $k$ has zero value") then interpretations $a_k, b_k, a_{3-k}, b_{3-k}$ should still the same in the next moment of time. Formulae B6 and B7 regulate the switching of truth values of $Q_i$ ($i \in \{1, \ldots L\}$).

Further, let the formula $\chi_0$ be conjunction of the following formulae:

- $Q_0 \wedge Same(d, a_1) \wedge Same(a_1, b_1) \wedge Same(b_1, a_2) \wedge Same(a_2, b_2)$ At the initial moment of time the constants $d, a_1, a_2, b_1, b_2$ have the same designation.
- $\bigcirc(Q_1 \wedge Same(a_1, a_2) \wedge Same(a_2, b_1) \wedge Same(b_1, b_2) \wedge \neg Same(a_1, d))$ At the next moment of time $Q_1$ holds and $d, a_1, a_2, b_1, b_2$ have the same designations, while $d$ has the different designation.
- $AlwaysNew(d) \wedge \Box(\bigwedge_{1 \leq i < j \leq L} \neg Same(e_i, e_j))$, stating that $d$ has different designations at different moments of time and $e_1, \ldots e_L$ all have different interpretations.

Finally, let $\chi^{\mathcal{M}}$ be $\bigwedge_{i=1}^{L-1}(\chi(c_k))$ where $\mathcal{M}$ is a Minsky machine defined by the sequence of instructions $c_1, \ldots, c_L$.

The formula $\chi_0 \wedge \chi^{\mathcal{M}}$ is intended to faithfully describe the computation of the machine $\mathcal{M}$ and the following lemma provides a formal justification for this.

**Lemma 1.** *A Minsky machine $\mathcal{M}$ produces an infinite run if, and only if, $\chi_0 \wedge \chi^{\mathcal{M}} \models \Box \neg Q_{stop}$.*

*Proof.*
$\boxed{\Rightarrow}$ Let a machine $\mathcal{M}$ produces an infinite run
$r^{\mathcal{M}} = (l_1, p_1^0, p_2^0), (l_2, p_1^1, p_2^1), \ldots (l_{j+1}, p_1^j, p_2^j), \ldots$, and a temporal structure $\mathfrak{M} = \langle D, \mathcal{I} \rangle$ is a model of $\chi_0 \wedge \chi^{\mathcal{M}}$. Straightforward induction on steps in $r^{\mathcal{M}}$ shows that, for all $j \geq 1$, the following relation between states of $\mathcal{M}$ and $\mathfrak{M}$ holds:

$l_j = l$ whenever $j \models Q_l$;
$p_1^j = |V_{a_1}^{j+1} \cap \overline{V_{b_1}^{j+1}}| = |V_{a_1}^{j+1}| - |V_{b_1}^{j+1}|$;
$p_2^j = |V_{a_2}^{j+1} \cap \overline{V_{b_2}^{j+1}}| = |V_{a_2}^{j+1}| - |V_{b_2}^{j+1}|$.

Since the run $r^{\mathcal{M}}$ is infinite we have $l_j \neq L$ for all $j \geq 1$, and therefore $j \models \neg Q_{stop}$ for all $j \geq 0$. Hence, $0 \models \Box \neg Q_{stop}$

$\boxed{\Leftarrow}$ By contraposition it is sufficient to show that if a machine $\mathcal{M}$ produces a finite run (halts) then $\chi_0 \wedge \chi^{\mathcal{M}} \wedge \Diamond Q_{stop}$ is satisfiable.

Let a machine, $\mathcal{M}$, halt and produce a finite run $r^{\mathcal{M}} = (l_1, p_1^0, p_2^0), \ldots (l_{s+1}, p_1^s, p_2^s)$, $s \geq 0$. The final executed instruction is the STOP instruction, so we have $l_{s+1} = L$. Now, we construct a temporal structure $\mathfrak{M}_n = \langle D, \mathcal{I}(n) \rangle$ as follows. We let the domain $D$ be a countable set. Then, for all $0 \leq j \leq s+1$, we ensure $j \models Q_l$ whenever $l_j = l$, and $j \models Q_{stop}$ for all $j > s+1$. Further we set $I(j)(d)$ (designations of $d$) to be different elements of the domain for all $j \geq 0$.

Further, we set
$0 \models \text{Same}(d, a_1) \wedge \text{Same}(a_1, b_1) \wedge \text{Same}(b_1, a_2) \wedge \text{Same}(a_2, b_2)$, and
$1 \models Q_1 \wedge Same(a_1, a_2) \wedge Same(a_2, b_1) \wedge Same(b_1, b_2) \wedge \neg Same(a_1, d)$.
Further define designations of $a_1, a_2, b_1, b_2$ for $2 \le j \le s$ inductively as follows:

- If the instruction with the label $l_j$ is of the first form (ADD) then define $I(j)(a_k) = I(m+1)(d)$, where $m$ is a such moment of time that $I(j-1)(a_k) = I(m)(d)$ and leave designations of the remaining constants the same as in $j-1$.
- If the instruction with label $l_j$ is of the second form (SUBTRACT) and $(j-1) \models \neg Same(a_k, b_k)$ then define $I(j)(b_k) = I(m+1)(d)$, where $m$ is a such moment of time that $I(j-1)(b_k) = I(m)(d)$ and leave designations of the remaining constants the same as in $j-1$.
- If the instruction with the label $l_j$ is of the second form (SUBTRACT) and $(j-1) \models Same(a_k, b_k)$ then leave designations of all constants ($a_k, b_k$, k=1,2) the same as in $j-1$.

Finally, assume designations of $a_1, b_1, a_2, b_2$ to be arbitrary for all $j > s$.

It is easily seen that this overall construction provides a model for $\chi_0 \wedge \chi^{\mathcal{M}}$ and since $l_{s+1} = L$ one also has $s \models Q_{stop}$. Thus, $\chi_0 \wedge \chi^{\mathcal{M}} \wedge \Diamond Q_{stop}$ is satisfied in $\mathfrak{M}$.

Now from Theorem 1 and Lemma 1 our main result follows:

**Theorem 2.** *The set of valid formulas of $LTL_{\lambda=}$ is not recursively enumerable.*

## 6 Conclusion

We have considered the extension $LTL_{\lambda=}$ of classical propositional temporal logic PTL and shown that the logic is suitable for specifications of dynamic systems using some resources, such as processes using memory locations or mobile agents occupying some sites. Despite its simplicity $LTL_{\lambda=}$ proved to be not recursively axiomatizable and, therefore, fully automated verification of $LTL_{\lambda=}$ specifications is not, in general, possible. Identification of decidable fragments of $LTL_{\lambda=}$ (if any) is an interesting problem for further research. We believe that undecidability holds also for the future time fragment of $LTL_{\lambda=}$. We leave detailed exposition of this case as well as the investigation of $LTL_\lambda$ (without equality) and $LTL_{\lambda(=)}$ with restrictions on the number of flexible constants to the future work.